\documentclass[12pt]{article}
\usepackage[dvips]{graphicx}
\usepackage{amssymb}
\usepackage{epsfig}

\catcode`@=11
\def\citer{\@ifnextchar
[{\@tempswatrue\@citexr}{\@tempswafalse\@citexr[]}}

\def\@citexr[#1]#2{\if@filesw\immediate\write\@auxout{\string\citation{#2}}\fi
  \def\@citea{}\@cite{\@for\@citeb:=#2\do
    {\@citea\def\@citea{--\penalty\@m}\@ifundefined
       {b@\@citeb}{{\bf ?}\@warning
       {Citation `\@citeb' on page \thepage \space undefined}}%
\hbox{\csname b@\@citeb\endcsname}}}{#1}}
\catcode`@=12

\newcommand{\ov}[1]{\overline{#1}}
\newcommand{\wt}[1]{\widetilde{#1}}
\newcommand{\ra}{\rightarrow}

\newcommand{\ee}{$e^+e^-$\ }

\newcommand{\epem}{e^+e^-}

\newcommand{\ttb}{t\ov{t}}
\newcommand{\bbb}{b\ov{b}}

\newcommand{\non}{\nonumber}
\newcommand{\lra}{\longrightarrow}
\newcommand{\beq}{\begin{eqnarray*}}
\newcommand{\eeq}{\end{eqnarray*}}
\newcommand{\SM}{SM~}
\newcommand{\SUSY}{SUSY~}
\newcommand{\MSSM}{MSSM~}

\newcommand{\tb}{{\rm tg} \beta}
\newcommand{\cc}{{\tilde{\chi}^+}}
\newcommand{\cn}{{\tilde{\chi}^0}}

\renewcommand{\thefootnote}{\fnsymbol{footnote}}

\newbox\mycount
\begin{document}

\begin{titlepage}

\begin{flushright}
DESY 99-178\\
\end{flushright}


\begin{center}

{\LARGE\sc Physics~with~an~$e^+e^-$~Linear~Collider\\[4mm]at~High~Luminosity%
\footnote{Based on lectures at the Carg\`{e}se 1999 Summer Institute,
          the Moscow 1999 Workshop on Quantum Field Theory,
          and the Lund 1999 Workshop on Future Electron-Positron Colliders.}}

\vspace{1.5cm}

{\large P.M.\ Zerwas \\
        Deutsches Elektronen--Synchrotron DESY \\[2mm]
        D--22603 Hamburg, FRG
}

\end{center}

\vspace{2.8cm}

\begin{abstract}
\noindent
The physics potential is briefly summarized 
for an $e^+e^-$ linear collider operating
at center--of--mass energies up to 
$\sqrt{s} = 1 \makebox{ TeV}$ and delivering  integrated luminosities
up to $\int {\cal{L}} = 0.5 \makebox{ ab}^{-1}$ in one to two  years.
This machine will allow us to perform precision studies 
of the top quark and the electroweak
gauge bosons at the per--mille level.
It will be an ideal instrument to investigate 
the properties of the Higgs boson and to establish
essential elements of the Higgs mechanism 
as the fundamental mechanism for breaking the electroweak symmetries.
In the area beyond the Standard Model, 
new particles and their interactions can be discovered and explored 
comprehensively.
In supersymmetric theories, the mechanism of the symmetry breaking 
can be investigated experimentally
and the underlying unified theory can be reconstructed. 
The high precision allows stable extrapolations 
up to scales near the Planck mass. 
\end{abstract}

\end{titlepage}

\section[]{Synopsis}

Essential elements of the fundamental constituents of matter and
their interactions have been discovered       in the
past three decades by operating \ee colliders.
A coherent picture of the
structure of matter has emerged, that is adequately described by
the Standard Model, in many of its facets
at a level of very high
accuracy. However, the Standard Model does not provide 
 a comprehensive theory of matter. Neither the fundamental
parameters, masses and couplings, nor the symmetry pattern can
be explained, but they are merely built into the
model. 
Moreover, gravity is not incorporated at the quantum level.
First steps to
solutions of these problems are associated with the unification of
the electroweak and the strong forces, and with the supersymmetric
extension of the model which provides
a bridge from the presently explored energy scales up to
scales close to the Planck mass.

Two strategies can be followed to enter into the area beyond
the Standard Model. 
{$(i)$} Properties of the particles
and forces within the Standard Model will be affected by new
energy scales. Precision studies of the top quark,
the electroweak gauge bosons 
and the Higgs boson
can thus reveal clues to the
physics beyond the Standard Model. 
{$(ii)$} Above the mass thresholds, new
phenomena can be searched for directly and
studied thoroughly so that the 
underlying basic theories can be reconstructed.\\

In this dual approach
a variety of fundamental problems still remain to be solved within
the \underline{Standard Model} \cite{1,1a},
demanding experiments at energies beyond the range of
existing accelerators.\\
\noindent
{$(a)$} The mass of the {\it top quark} is much larger
than the masses  of the electroweak gauge bosons. 
Understanding the r\^ole of this
particle in Nature is therefore an important goal for the future.
In the $\ttb$ threshold region of \ee
collisions the top quark mass can be measured to
an accuracy better than 200 MeV. 
This is a desirable goal since a future
theory of flavor dynamics will lead to   relations among the
lepton/quark masses and mixing angles in which
the heavy top quark is
expected to play a key r\^ole. 
In addition, stringent tests in
the electroweak and Higgs sector of the Standard Model can be carried out when
the top mass is known very accurately. Helicity analyses of the $\ttb$
production vertex and the $t$ decay vertex will determine the magnetic
dipole moments of the top quark and the chirality of the ($tb$)
decay current at the per--cent level. 
Bounds on the ${\cal {CP}}$
violating electric dipole moments
of the $t$ quark can be set to $10^{-18}$\,e\,cm.\\
\noindent
{$(b)$} Studying the dynamics of the {\it electroweak gauge bosons}
is another important task at high energy \ee colliders. The form
and the strength of the triple and quartic couplings of these particles
are uniquely predicted by the  non--abelian gauge symmetry
of the theory,
defining the electroweak
charges, the magnetic dipole moments and the electric quadrupole
moments of the $W^\pm$ bosons in the static limit. 
Tests of these fundamental symmetry concepts can be performed 
at an accuracy of $10^{-3}$ down to $10^{-4}$.\\
\noindent
{$(c)$} A high--luminosity \ee collider with
an energy between 300 and 500\,GeV will be an ideal
instrument to search for {\it Higgs particles} \cite{1b}
throughout the mass
range characterized by the scale of electroweak symmetry breaking,
and to investigate their properties.
The intermediate Higgs mass range below $\sim 200$ GeV is 
the theoretically                    preferred region.
 In this scenario
Higgs particles remain weakly interacting up to the scale of
grand unification, thus providing a basis for the renormalization
of the electroweak mixing angle $\sin^2 \theta_W$ from the GUT
symmetry value 3/8 down to the experimentally
observed value close to 0.2.
Once the Higgs particle is found, its properties can
be studied thoroughly, the external quantum numbers
${\cal J}^{\cal{PC}}$ and the
Higgs couplings, including the self--couplings of the particle.
The measurements of these couplings are the necessary ingredient
to establish the Higgs mechanism {\it sui generis} experimentally.\\

Even though many facets of the Standard Model are experimentally
supported at a level of  very high accuracy,
extensions should nevertheless be anticipated
as argued before. The next generation of accelerators can shed
light on three domains in the area beyond the Standard Model.

\vspace*{1mm}
The \underline{Grand Unification} of the gauge symmetries \cite{2} 
suggests itself quite naturally. 
This idea can be realized in different
scenarios some of which
predict new vector bosons and a plethora of new fermions.
Mass scales of these novel particles could be as low as a few
hundred GeV.

\vspace*{1mm}
A very important theoretical extension of the Standard Model,
which is interrelated with
the unification of the gauge symmetries, is
\underline{Supersymmetry} \cite{3}. 
This novel symmetry concept unifies matter and
forces by pairing the associated fermionic and bosonic
particles in common multiplets. Several arguments strongly
support the hypothesis that this symmetry is
realized in Nature indeed. 
{$(i)$} Supersymmetry stabilizes
light masses of Higgs particles in the context of very
high energy scales as demanded by grand unified theories.
{$(ii)$} Supersymmetry may generate
the Higgs mechanism itself by inducing the radiative symmetry breaking of
${\rm SU(2)_L \times U(1)_Y}$
while leaving ${\rm U(1)_{EM}}$ and ${\rm SU(3)_c}$ unbroken
for a top quark mass between 100 and 200\,GeV.
{$(iii)$} This symmetry picture is also supported strongly by the
successful prediction of the electroweak mixing angle in the
minimal version of the theory.  The particle spectrum in
this theory drives the evolution of the electroweak mixing
angle  from the
GUT value 3/8 down to $\sin^2\theta_W$ = 0.2336 $\pm$ 0.0017, 
within a margin
 $\sim$ 0.002 to
the  experimental value $\sin^2 \theta_W^{exp}$ =
0.2316 $\pm$ 0.0002.

A spectrum of
several neutral and charged
Higgs bosons is predicted in supersymmetric theories.
The mass of the lightest Higgs boson is 
less than $\sim 150$ GeV
in nearly all scenarios
while the heavy Higgs particles have masses of the order
of the electroweak symmetry breaking scale.  Many other
novel particles are predicted in supersymmetric theories.
The scalar partners of the leptons could have masses in the
range of $\sim 200$ GeV whereas squarks are expected to be
considerably heavier. 
The lightest supersymmetric states are likely to be
non--colored gaugino/higgsino states with masses
possibly in the 100 GeV range. Searching for these supersymmetric
particles will be one of the most important tasks at future
\ee colliders. 
Moreover, the high accuracy which can be achieved
when masses and couplings are measured, will allow
us to 
determine the mechanism of supersymmetry breaking and to
extrapolate the basic parameters of the theory so that 
the key elements of the underlying grand unified
theories at scales, potentially close to the Planck scale, 
can be reconstructed.

\vspace*{1mm}
In the alternative scenario of heavy or no fundamental
Higgs bosons, \underline{new strong inter-} \underline{actions} between
electroweak bosons would be observed, characterized by a
scale of order 1 TeV at which the electroweak symmetries would
be broken strongly \cite{5A}.
This scenario could be analyzed by studying the elastic
scattering of $W$ bosons at high energies. New resonances 
would be formed, the properties of which would uncover the
underlying microscopic interactions.\\

While new high--mass vector bosons and particles carrying color
quantum numbers can be searched for very efficiently at hadron
colliders, \ee colliders provide in many ways
unique opportunities to discover
and explore non--colored particles.
This is most obvious  in supersymmetric theories. Combining
LEP2 analyses with future searches at the Tevatron and the LHC, the light and
heavy Higgs bosons can  be found individually only
in part of the supersymmetry parameter space.
Squarks and gluinos can be
searched for very efficiently at the LHC. 
Yet precision studies of their properties
are possible only in part of the parameter space.
Similarly non-colored supersymmetric particles;
a model--independent analysis of gauginos/higgsinos and scalar sleptons
can only be carried out at
\ee colliders with well--defined kinematics at
the level of the subprocesses.
The detailed knowledge of all the properties of
the colored and non--colored supersymmetric
states will reveal the mechanism of supersymmetry breaking and
the structure of the underlying theory.

Thus, the physics progamme of $e^+e^-$ linear colliders
is in many
aspects complementary to the programme of the $pp$ collider LHC.
The high accuracy which can be achieved at $e^+e^-$ colliders in
exploring the properties of the top quark, electroweak gauge
bosons, Higgs particles and supersymmetric particles will enable us
to cover the energy range above the existing machines up to the TeV
region in a conclusive form, eventually providing us with essential
clues to the basic structure of matter and the laws of physics.

\vspace*{3mm}
The discussion will focus on the physics program at an $e^+e^-$ linear 
collider
operating at center--of--mass energies above LEP2 up to about 1\,TeV.
Primarily  high--luminosity runs, collecting integrated 
luminosities
up to 0.5\,ab$^{-1}$ in one to two years of operation, will be described.
Also the results expected from high--luminosity runs at low energies on the
$Z$ resonance, the GigaZ mode, and near the $WW$ threshold will be 
summarized.
Electrons and positrons will in general be assumed polarized to 80\%
and 60\%, respectively.
Specific problems which can be solved in 
$e^-e^-$, e$\gamma$ and $\gamma\gamma$ modes
of the linear collider will be addressed in the appropriate context.

This summary report is built on the general linear collider review 
of Ref.\cite{4}.
Other material can be found in Refs.\cite{5},
experimental aspects  particularly in Ref.\cite{6}.
For recent summaries of the LHC physics and $\mu\mu$ physics programs
see  Refs.\cite{6a} and \cite{6b}, respectively.

\vspace*{0.5cm}

\section{Top Quark Physics}

Top quarks are the heaviest matter particles in the 3--family
Standard Model, introduced to incorporate
${\cal {CP}}$ violation in the
left--handed charged current sector.
They may therefore hold the key for aspects of the physics
beyond the Standard Model at high--energy scales.
Examples in which the large top mass is crucial, are
multi--Higgs doublet models, models of dynamical symmetry breaking,
compositeness and supersymmetry.
Strong indirect evidence for the existence of top quarks,
based on the well established gauge symmetry pattern of
the Standard Model, had been
accumulated quite early.
By evaluating the high--precision electroweak data, the value
of the top quark mass was estimated to be
$m_t = 180 \pm 14$ GeV. 
Top quarks have recently
been observed directly by the two Tevatron
experiments \cite{7},
corresponding to a mass of $m_t = 174.3 \pm 5.1 \mbox{ GeV}$
which is in striking agreement with the result of the electroweak
 analysis.

\subsection[]{The Profile of the Top Quark}

For a top mass larger than the $W$ mass, the channel
\[
t \ra b + W^+
\]
is the dominant decay mode. For m$_t \sim $ 175\,GeV
the width of the top quark, $\Gamma_t \sim $ 1.4\,GeV, is so large
compared with the scale $\Lambda$ of the strong interactions
that this quark can be treated as a bare quantum which is not dressed 
by non--perturbative strong interactions \cite{8}.\\[2mm]
\noindent
\underline{Chirality of the $(tb)$ decay current:}
The precise determination of the weak isospin quantum
numbers does not allow for large deviations of the
$(tb)$ decay current from the left--handed \SM\  prescription.
Nevertheless, since $V+A$ admixtures may grow with the masses of the
quarks involved [through mixing with heavy mirror quarks, for
instance], it is necessary to check the chirality of the decay
current directly. The $l^+$ energy distribution in
the semileptonic decay chain $t \ra W^+ \ra l^+$
depends
on the chirality of the current. Any deviation from the
standard $V-A$ current would lead to a stiffening of the
spectrum and, in particular, to a non-zero value at the upper
end--point of the energy distribution. A sensitivity of about 5\%
to a possible $V + A$ admixture can be reached experimentally
\cite{9}.\\[2mm]
\noindent
\underline{Non--standard top decays} could occur in supersymmetric extensions
of the Standard Model: top decays into charged Higgs bosons and/or
top decays to stop particles,
$t \ra b+H^+$
and
$t \ra \tilde{t} + \tilde{\chi}^o_1$.
If kinematically allowed, branching ratios of order 10\% are expected
in both cases so that these decay modes could
be observed easily.
Decays with signatures as clean as $t \ra c\gamma,cZ, cH$ may be detected for 
branching ratios of order $10^{-4}$ and less.\\

The main production mechanism for top quarks in \ee collisions is the
annihilation channel \cite{10},
\[
e^+e^- \stackrel{\gamma, Z}{\longrightarrow} t \ov{t}
\]
For $m_t  \sim 175$ GeV, the maximum of the cross section
$\sigma (t \ov{t}) \sim 800$ fb
is reached about 30\,GeV above the threshold,
giving rise to a million top quarks in two years of collider operation.
If the scale of new areas beyond the Standard Model is
much larger than the collider energy,
the electroweak production currents can globally be described by form factors
which reduce to anomalous $Z$ charges, anomalous magnetic dipole moments
and electric dipole moments.\\[3mm]
\noindent
\underline{Magnetic dipole moments:}
If the electrons in
$e^+e^- \ra t\ov{t}$ are
left--handedly polarized, the top quarks are produced preferentially
as left--handed particles in the forward direction while only a small
fraction is produced as right--handed particles in the backward
direction \cite{11}, so that
 the backward
direction is most sensitive to small anomalous magnetic moments of
the top quarks. 
The anomalous
magnetic moments can be bounded to the percent level by
measuring the angular dependence of the $t$ quark cross section
in this region.\\[3mm]
\noindent
\underline{Electric dipole moments:}
These moments are generated by ${\cal {CP}}$--non
invariant interactions. Non--zero values of the moments can be
detected by means of  non--vanishing expectation values of ${\cal {CP}}$--odd
momentum tensors such as $T_{ij} = (q_+ - q_-)_i (q_+ \times
q_-)_j$ with $q_{\pm}$ being the unit momentum vectors of the
$W$~decay leptons. Sensitivity limits to $\gamma, Z$ electric
dipole moments of $\lesssim 10^{-18}$ e\,cm can be reached
\cite{12} for an
integrated luminosity of $\int{\cal{L}} = 100 \mbox{ fb}^{-1}$
at $\sqrt{s} = 500$\,GeV.

\subsection[]{The Top-Quark Mass}

Quark--antiquark production near the threshold in $e^+e^-$
collisions is of exceptional interest. 
The long time which
the particles stay close together at low velocities,
allows the strong interactions to
build up rich structures of bound states and resonances.
This picture would have applied
to top quarks up to the mass range of $\sim
130$ GeV. Beyond this value,  the picture changes 
quite dramatically as a result of  the rapid top decay:
The decay time of the states
becomes shorter than the revolution time of the constituents so
that toponium resonances cannot be formed any more \cite{8}. 
For a while, however, remnants of the $1S$ state give rise to a peak in the
excitation curve, yet it disappears for top masses in excess of 180\,GeV. 
Nevertheless, across this range the resonance remnants induce
a steep rise of the cross section near the threshold.

\begin{figure}[t]
\centerline{\includegraphics[width=8cm,angle=90]{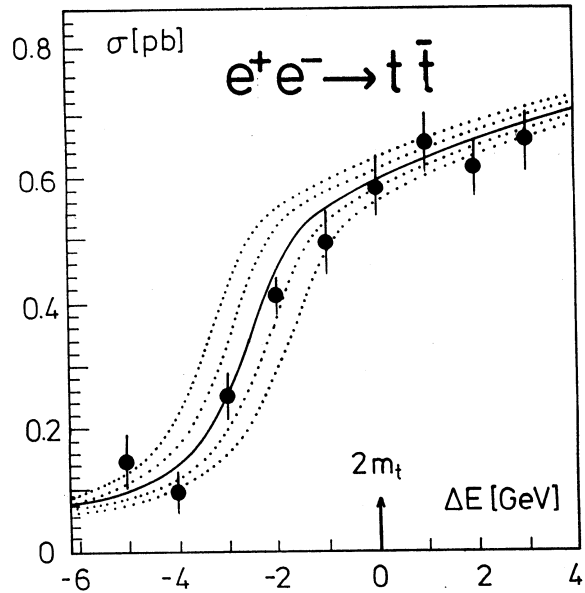}}
\caption{\label{fig1}\it The excitation curve for top production near the threshold; Ref.\cite{16a}.}
\end{figure}

Since the rapid top
decay restricts the interaction region to small distances, the
excitation curve can be predicted in perturbative QCD \cite{16},
based essentially on the Coulombic interquark potential
$ V(R) = - 4/3 \times \alpha_s(R)/R$. 
The cross section
is built up by the superposition of all $nS (t \bar{t})$ states.
The form and the height of the excitation curve are very sensitive
to the mass of the top quark,
cf Fig.~\ref{fig1}.

Detailed experimental simulations predict the following sensitivity
to the top mass \cite{19} near $m_t \sim 175$\,GeV:
\begin{eqnarray*}
\delta m_t & \lesssim & 200 \ \mbox{MeV}
\end{eqnarray*}
for an integrated luminosity of
$\int{\cal{L}} = 50 \mbox{ fb}^{-1}$, including
remnant uncertainties due to higher--order QCD corrections
and experimental errors contributing at approximately equal strength.
At proton colliders a sensitivity of about 1 to 2\,GeV has been predicted
 for the top mass, based on the
reconstruction of top quarks from jet and lepton final states \cite{20A}.
Thus, \ee colliders will improve the
measurement of the top quark mass by
at least an order of magnitude.

\vspace{0.5cm}

\section[]{Electroweak Gauge Bosons}

\subsection[]{Standard $W$, $Z$ Bosons}

The fundamental electroweak and strong forces appear to be of gauge
theoretical origin. This is one of the outstanding results
of theoretical and experimental analyses in the past three decades.
However, little direct evidence has been accumulated so far
for the non--Abelian nature of the forces in the
electroweak $W^\pm, Z, \gamma$ sector.
Since deviations from the gauge
symmetries manifest themselves in experimental observables
with coefficients $(\beta\gamma)^2$, 
high energies will allow stringent direct tests of
the self--couplings of the electroweak gauge bosons. 

The gauge symmetries of the Standard Model determine the form and
the strength
of the self--interactions of the electroweak bosons, triple couplings
$WW\gamma,WWZ$ and quartic couplings.
Deviations from the gauge symmetric form of these vertices
could be expected in more general scenarios \cite{20}. 
In models in which $W,Z$ bosons are generated dynamically 
and interact strongly with each other at high scales $\Lambda_\ast$,
corrections could alter the vertices to order $(M_W/ \Lambda_\ast)^2$
and induce new types of couplings.

While the experimental analyses of the
self--couplings of the electroweak bosons can be carried out at collider
energies of 500 GeV with high accuracy, $WW$ scattering \cite{20a}
can only be studied at energies in the TeV range. 
This is an important process 
which must be investigated very thoroughly
if light Higgs particles do not exist and
$W$ bosons become strongly interacting particles at high energies.

The properties of the $Z$ boson have been studied at LEP and SLC
with very high accuracy. By operating TESLA at low energies
on the $Z$ resonance in the GigaZ mode and near the $WW$ threshold,
the measurement of fundamental electroweak parameters, the
electroweak 
mixing angle 
and $W$ mass can be improved by yet an order of magnitude,
allowing to test electroweak symmetry breaking stringently at
the quantum level.

\paragraph{\underline{The GigaZ Mode:}}
By building a bypass for the transport of electron and positron bunches,
high luminosity can also be reached at low energies in linear colliders. 
On the $Z$ resonance, 
an ensemble of $10^9$ events, 1\,GigaZ, can be generated in a year,
expanding the LEP sample by two orders of magnitude.
With both electron and positron beams longitudinal polarized, 
the electroweak mixing angle can be determined very accurately
by measuring the left--right asymmetry 
$A_{LR} = 2\,(1-4 \sin^2 \theta_W) \, / \,[1+(1-4 \sin^2 \theta_W)^2]$,  
Ref.~\cite{20b}:
\begin{eqnarray*}
\delta \sin^2 \theta_W \approx 10^{-5}
\end{eqnarray*}
Similarly the measurement of the $W$ mass can be improved to
\begin{eqnarray*}
\delta M_W \approx 6\,\makebox{MeV}
\end{eqnarray*}
by scanning the threshold region, Ref.~\cite{20c}.

Based on these two measurements, a variety of high--precision tests
can be performed in the electroweak sector. 
Extracting the Higgs mass from the
electroweak observables is particularly interesting. 
From the $\rho$--parameter \cite{20e1}
this mass can be predicted
to an accuracy of about 6\%, improving LHC based predictions by
almost an order of magnitude \cite{20d}. 
Comparing this prediction with the direct measurement of the Higgs mass,
quantum fluctuations can be tested stringently in a spontaneously broken gauge theory.

\paragraph{\underline{The Triple Gauge-Boson Couplings:}}
The couplings $W^+W^-\gamma$ and $W^+W^-Z$ are in general described
each by seven parameters. Assuming 
${\cal C},{\cal P}$ and ${\cal T}$ 
invariance in the pure electroweak boson sector, the number of
parameters can be reduced to three,
\begin{eqnarray*}
{\cal L}_k /ig_k &=&  g_k^1 W^*_{\mu \nu} W_\mu A_\nu + 
{\rm h.c.} \ + \   \kappa_k W^*_\mu W_\nu  F_{\mu \nu} \ + 
\  \frac{\lambda_k}{M_W^2} 
W^*_{\rho \mu} W_{\mu \nu} F_{\nu \rho}
\end{eqnarray*}
with $g_\gamma = e$ and $g_Z = e\cot\theta_W$ for $k=\gamma, Z$. 
The $\kappa = 1 + \Delta \kappa$ and the $\lambda$ parameters
can be identified with the $\gamma, Z$ charges of the $W$
bosons and the
related magnetic dipole moments and electric quadrupole moments:
\begin{eqnarray*}
\mu_\gamma \quad = 
&\frac{\displaystyle e}{\displaystyle 2M_W}\Big[ 2+\Delta\kappa_\gamma +\lambda_\gamma \Big]
&{\rm and} \quad \gamma \ra Z \\
Q_\gamma   \quad = 
&-\frac{\displaystyle e}{\displaystyle M_W^2} \Big[ 1+\Delta\kappa_\gamma - \lambda_\gamma \Big]
&{\rm and} \quad \gamma \ra Z
\end{eqnarray*}
The gauge symmetries of the \SM demand $\kappa = 1$ and $\lambda= 0$.
The magnetic dipole and the electric quadrupole moments
can be measured {\it directly} in the production of
 $W \gamma$ and $WZ$ pairs
at $p\overline{p} / pp$ colliders and $WW$ pairs at \ee
and $\gamma \gamma$ colliders.

Detailed experimental analyses  have been carried out for the reaction
$\epem \rightarrow W^+ W^- \ra (l \nu_e)(q\overline{q}\,')$.
The bounds on $\Delta \kappa, \lambda$ which
can be obtained at \ee colliders of 500\,GeV \cite{20e}
are significantly better than the bounds expected from the LHC: 
\begin{eqnarray*}
  & \Delta g_1^Z     = 2.5 \times 10^{-3}\\
\Delta \kappa_\gamma = 4.8 \times 10^{-4} \qquad & \Delta \kappa_Z  = 7.9 \times 10^{-4} \\
\lambda_\gamma       = 7.2 \times 10^{-4} \qquad &  \lambda_Z = 6.5 \times 10^{-4}                      
\end{eqnarray*}
Moreover, they
improve at 1\,TeV by nearly an order of magnitude. 
The scales $\Lambda_\ast$ which
can be probed, extend far beyond the energy scales which are accessible
directly.

\paragraph{\underline{Strongly Interacting $W$, $Z$ Bosons}}
If the scenario in which $W$, $Z$, Higgs bosons are weakly interacting
up to the
GUT scale  is not realized
in Nature, the alternative scenario is a strongly interacting $W$,
$Z$ sector. 
Without a light Higgs boson with a mass
of less than about 1\,TeV, the
electroweak bosons must become strongly interacting particles at energies of
about 1.2\,TeV to comply with the requirements of unitarity
for the $W_LW_L$ scattering amplitudes.
By absorbing the Goldstone particles associated
with the spontaneous symmetry breaking of 
the new strong interactions,
the
longitudinal degrees of freedom for the massive vector bosons
may be built up, as realized in
technicolor type theories, for instance. 
In such scenarios, novel  resonances are predicted     in the 
${\cal O}$(1\,TeV) energy range which can be 
generated in $W_LW_L$ collisions.

In scenarios of strongly interacting vector bosons, $W_LW_L$ scattering
must be studied at energies of order 1 TeV which requires the
highest energies possible in $e^{\pm}e^-$ colliders.
(Quasi)elastic
$WW$ scattering can be investigated by using $W$ bosons radiated off the
electron and positron beams, $ ee \ra \nu \nu WW$,
or by exploiting final-state interactions in the
\ee annihilation to $W$ pairs, $e^+e^- \ra W^+W^-$.
All possible (isospin, angular momentum) combinations
in $WW$ scattering amplitudes
$a_{IJ}$ can be realized in the first process.
The cross sections however
are small as long as no resonances are formed. 

Building up the electroweak vector boson masses by the interactions
of the gauge fields with the Goldstone bosons
associated with the spontaneous symmetry
breaking of the underlying strong--interaction theory,
the longitudinal degrees of freedom of the
vector bosons can be identified at high energies with the Goldstone
bosons themselves
as a result of the equivalence theorem. 
In analogy to the $\pi\pi$ low--energy theorems, the first terms in the energy
expansion of the $WW$ scattering amplitudes \cite{20f}
are determined independent of dynamical details: 
\begin{eqnarray*}
a_{00} = +\frac{6}{96\pi v^2} & \qquad & a_{20} = -\frac{2}{96\pi v^2}\\[2mm]
a_{11} = +\frac{1}{96\pi v^2} &
\end{eqnarray*}
These fundamental scattering amplitudes in the threshold region
of the strong $WW$ interactions can be tested \cite{20g} to an accuracy
of about 15\% at a high--luminosity collider of 1\,TeV; new
strong-interaction scales extending up to about 3\,TeV can be
probed in these experiments, thus covering energy scales
up to the formation of novel resonances.

The attractive $I=0$ and $I=1$ channels may form
Higgs and $\rho$--type resonances at high energies. 
The formation of resonances would 
lead to spectacular phenomena in $WW$ collisions \cite{20h}.

Similar phenomena would also be observed 
as rescattering effects
in the cross section
$\sigma(\epem \ra W^+W^-)$ for $W$--pair production.
$(I,J) = (1,1)$ resonance effects would be noticeable at
$\sqrt{s} = 1$ TeV up to resonance masses of about 5 TeV in the angular
distributions of the $W$ decay final states \cite{24}.

\subsection[]{Extended Gauge Theories}

The gauge symmetry of the Standard Model, ${\rm SU(3) \times
SU(2) \times U(1)}$, is
widely believed not to be
the {\it ultima ratio}.
The \SM does not unify the electroweak and strong forces
since the coupling constants of these interactions are
different and appear to be independent.  
However, one should expect that in a more
fundamental theory  the three forces are described within
a single gauge group and, hence, with only one coupling constant
at high energy scales.
This grand unified theory will be based on a gauge group containing
${\rm SU(3) \times SU(2) \times U(1)}$ as a
subgroup and it will be reduced to this symmetry at low energies.

Two predictions of grand unified theories may have interesting
consequences in the energy range of a few  hundred GeV \cite{25}:

{\bf{(a)}}
The unified symmetry group must be spontaneously broken at the
unification scale
$\Lambda_{\rm GUT} \stackrel{<}{\sim}  10^{16}$\,GeV
in order to be compatible
with the experimental bounds on the proton lifetime.  
However, the breaking to the \SM group may occur in several steps and
some subgroups may remain unbroken  down to a scale of order 1\,TeV.
In this case
the surviving group factors allow for new \underline{gauge bosons} with
masses not far above the scale of electroweak symmetry breaking.
Besides $\rm SU(5)$, two other unification groups have received much
attention: In $\rm SO(10)$ three new gauge bosons
$W^\pm_R, Z_R$ are predicted, while in
E(6) a light neutral $Z'$ boson may
exist in the TeV range.

The virtual effects of a new $Z_R/Z'$ boson associated with the most general
effective theories which arise from breaking E(6) $\ra {\rm SU(3)
\times SU(2) \times U(1) \times
U(1)_{Y'}}$ and ${\rm SO(10) \ra \ SU(2)_L \times SU(2)_R \times U(1)}$,
have been investigated in Ref.~\cite{25a}.
Assuming  the $Z_R/Z'$ bosons to be heavier than the available
c.m. energy, the propagator effects on various observables of the
process
\begin{eqnarray*}
\epem \stackrel{V}{\longrightarrow} f\bar{f} \, : \quad V = \gamma,Z ~\makebox{and}~ Z_R/Z'
\end{eqnarray*}
have been studied in detail. 
As shown in Table~\ref{tab1}, the effects of new vector bosons
can be probed for masses up to 5\,TeV at a 500\,GeV collider.
While they may be produced directly up to about 5\,TeV
at the LHC, experiments at the \ee collider will measure the
couplings of the vector bosons to fermions very precisely,
thus identifying the physical nature of the new bosons.
Masses up to 10\,TeV and 50\,TeV can be probed in \ee colliders
operating at 800\,GeV and 5\,TeV, respectively.
These two windows extend to much higher scales than
the discovery limits anticipated at LHC.

\begin{table}
\begin{center}
\begin{tabular}{|r||r|r|}
\hline
\rule[-3mm]{0mm}{9mm} $\sqrt{s}$\,\,\,&  SO(10)          &  E(6)\,\,        \\
\hline
\rule[-0mm]{0mm}{6mm} 500\,GeV        &  6\,TeV          &  5--7\,TeV       \\
                      800\,GeV        &  10\,TeV         &  8--11\,TeV      \\
\rule[-3mm]{0mm}{8mm} 5\,TeV          &  $\sim$\,50\,TeV &  $\sim$\,50\,TeV \\
\hline
\end{tabular}
\caption{\label{tab1}{\it Sensitivity limits of $Z_R$ masses in SO(10)
and $Z'$ masses in E(6) at $e^+e^-$ linear colliders in the TeV range.}}
\end{center}
\end{table}

{\bf{(b)}} The grand unification  groups incorporate extended
fermion representations
in which a complete generation of \SM quarks and leptons can be
naturally embedded. These representations accommodate a variety
of additional new \underline{fermions}. It is conceivable that
the new fermions  acquire masses not much
larger than the Fermi scale. 
This is necessary if the predicted new gauge bosons are relatively light.  
SO(10) is the simplest group in which the 15 members of
each \SM generation of fermions can be embedded into a single
multiplet.  
This representation has dimension {\bf 16} and
contains a right--handed neutrino. The group E(6) contains
$\rm {SU(5)}$ and $\rm {SO(10)}$ as subgroups.
In E(6), each quark--lepton
generation belongs to a representation of dimension
{\bf 27}. To complete this representation, twelve new fields are needed
in addition to the \SM fermion
fields. In  each family the spectrum includes two additional
isodoublets of leptons, two isosinglet  neutrinos and an isosinglet
quark with charge $-1/3$.

If the new particles have non--zero electromagnetic and  weak charges,
and if their masses are smaller than the beam energy of the
$\epem$ collider, they  can be pair produced.
In general, the production processes are built  up by
a superposition of s--channel
$\gamma$ and $Z$ exchanges, but additional
contributions could
come from the extra neutral bosons if their masses are not much larger
than the c.m.~energy \cite{25}.
The cross sections are large,
of the order of the  point--like QED cross section.
This leads to samples of several thousands of events.
Fermion mixing, if large enough, gives rise to
additional production mechanisms for the new
fermions: single production in association with the light partners.
In this case, masses very close to the total energy of the $\epem$
collider can be reached.

\section{The Higgs Mechanism}

\subsection{Basis}
The Higgs mechanism is the cornerstone in the electroweak sector
of the Standard Model. The fundamental \SM particles, leptons,
quarks and weak gauge bosons, acquire masses by means of  the
interaction with
a scalar field. To accommodate the well--established electromagnetic and
weak phenomena,
the Higgs mechanism requires the existence of at
least one weak iso--doublet scalar field. 
After absorbing three Goldstone modes
to build up the longitudinal polarization
states of the $W^\pm,Z$ bosons, one degree of freedom is
left--over, corresponding to a real scalar particle. 

Three steps are necessary to establish experimentally
the Higgs mechanism {\it sui generis} 
 as the mechanism for generating the masses of
the fundamental \SM particles:
\begin{description}
\item[(i)\hspace*{2.5mm}]  The Higgs boson must be 
discovered -- the {\it experimentum crucis};
\item[(ii)\hspace*{1.5mm}] The couplings of the Higgs particle
with gauge bosons and fermions must be proven to 
 increase with their masses;
\item[(iii)] The Higgs potential generating the non--zero Higgs field
in the vacuum and breaking the electroweak symmetry in the scalar sector
must be reconstructed by determining the Higgs self--couplings.
\end{description}

The only unknown parameter in the \SM Higgs sector is the mass of the
Higgs particle. Constraints on the mass can, however, be
derived from the upper scale $\Lambda_{\ast}$ of
the energy range in which the model is assumed to
be valid before the particles become strongly interacting
and new dynamical phenomena emerge \cite{29}.
Increasing the energy scale, 
the quartic self--coupling of the Higgs field 
grows 
for large values indefinitely.
If the Higgs
mass is small, the energy cut--off $\Lambda_{\ast}$ is large at which
the coupling
grows beyond any bound; conversely, if the
Higgs mass is large, the cut--off $\Lambda_{\ast}$ is small. The condition
$M_H < \Lambda_{\ast}$ sets an upper limit on the Higgs mass in the Standard
Model. Detailed
analyses lead to an estimate
of about 700 GeV for the upper limit on $M_H$.
If the Higgs
mass is less than 180 to 200\,GeV, the Standard Model can be extended
up to the
GUT scale $\Lambda_{{\rm GUT}} \sim 10^{16}$ GeV, while all particles
remain weakly interacting. The hypothesis that the interactions
between
$W,Z$ bosons
and Higgs particles remain weak up to the GUT scale,
plays a key r\^ole in deriving the experimental value of the
electroweak mixing parameter $\sin^2 \theta_W$ from
grand unified theories.
From this hypothesis and the additional
requirement of vacuum stability,
upper and lower bounds on
the Higgs mass can be derived. Based on these arguments, the \SM
Higgs mass should
be expected in the mass window $130 < M_H < 180$\,GeV for a top mass
value of about 175\,GeV.

Several            channels can be exploited to search for
Higgs particles in the
Higgs--strahlung and fusion processes of \ee colliders \citer{30,30b}.
In the Higgs--strahlung process \ee $\ra ZH$,
missing--mass techniques can be used
in events with leptonic $Z$ decays or
the Higgs particle
may be reconstructed in $H \ra b \bar{b}, WW$ directly.
The $WW$ fusion process \ee $\ra \bar{\nu}_e \nu_e H$ requires the
reconstruction of the Higgs particle.

Once the Higgs boson is found at LEP, Tevatron or LHC, it will be very 
important
to explore its properties at the \ee linear collider to establish 
the Higgs mechanism
 experimentally.
This is possible with high precision
in the clean environment of $\epem$ colliders
in which 
at high luminosity
a large ensemble of order $10^5$ Higgs bosons 
can be generated nearly background--free.
The zero--spin of
the Higgs   particle is reflected in the angular distribution of the
Higgs--strahlung process
which must approach the $\sin ^2\theta$ law asymptotically . 
The strength of the couplings to $Z$ and $W$ bosons
is reflected in
the magnitude of the \ee production cross sections.
The strength of the  couplings to fermions can be measured in the
decay branching ratios
and the Higgs bremsstrahlung off top quarks.
Double Higgs--strahlung can be exploited to measure the trilinear
 Higgs self--coupling.

From the preceding discussion we conclude that an $\epem$ linear
collider  with   energies in the range of 300 to 500 GeV  
and high luminosity is the ideal
instrument to investigate the Higgs mechanism in the intermediate mass
range which, {\it a priori}, may be considered the theoretically preferred
part
in the entire range  of  possible Higgs mass values. \\

\subsection{The Higgs Particle in the Standard Model}

The profile of the \SM Higgs particle is completely
determined if the Higgs mass is fixed.
For Higgs particles in the intermediate mass range
$M_Z \leq M_H \leq 2M_Z$
the main decay modes are decays into $b \bar{b}$ pairs and $WW,ZZ$ pairs
with one of the two gauge bosons being virtual
below the threshold \cite{31}.
Above the $WW$ threshold, the Higgs particles decay almost
exclusively into these channels, except in the mass range near
the $t \overline{t}$ decay threshold.
Below 140 GeV, the decays $ H \ra \tau^+ \tau^-, c \bar{c}$ and
$gg$ are also important       besides the dominating $b \bar{b}$
channel. Up to masses of 140 GeV, the Higgs  particle is very narrow,
$\Gamma(H) \leq 10$ MeV. After opening the [virtual] gauge boson channels,
the state becomes rapidly wider, the width
reaching $\sim$ 1 GeV at the $ZZ$ threshold.
The width cannot be measured directly in the intermediate mass range.
Only above $M_H \geq 200$ GeV it becomes wide enough to be resolved
experimentally.

The main production mechanisms for Higgs particles in $\epem$ collisions
are Higgs--strahlung off the $Z$ boson line \cite{33} and the $WW$ fusion
process \cite{34},
\begin{eqnarray*}
(a) \quad \epem & \stackrel{Z}{\longrightarrow}  &  Z+H  \\
(b) \quad \epem & \stackrel{WW}{\longrightarrow} &  \bar{\nu}_e \ \nu_e \ +H
\end{eqnarray*}
With rising energy the
Higgs--strahlung cross section scales $\sim\,\alpha^2_w/s$
while the fusion
cross sections increase logarithmically
$\sim\,\alpha^3_w M_W^{-2} \mbox{log } s/M_H^2$, becoming dominant above 500\,GeV:
\begin{eqnarray*}
\sigma(e^+e^- \ra ZH) & \ra & \frac{G_F^2 M_Z^4}{96\pi s}[1+(1-4 \sin^2\theta_W)^2] \\[3mm]
\sigma(e^+e^- \ra \bar\nu\nu H) & \ra & \frac{G_F^3 M_W^4}{4 \sqrt{2} \pi^3} \log \frac{s}{M_H^2}
\end{eqnarray*}
As a general rule, the cross sections and rates 
[about $10^5$ events]       
are sufficiently large to detect
Higgs particles with masses up to 70\% of the total \ee c.m. energy.

\begin{figure}[t]
\centerline{\includegraphics[width=8cm]{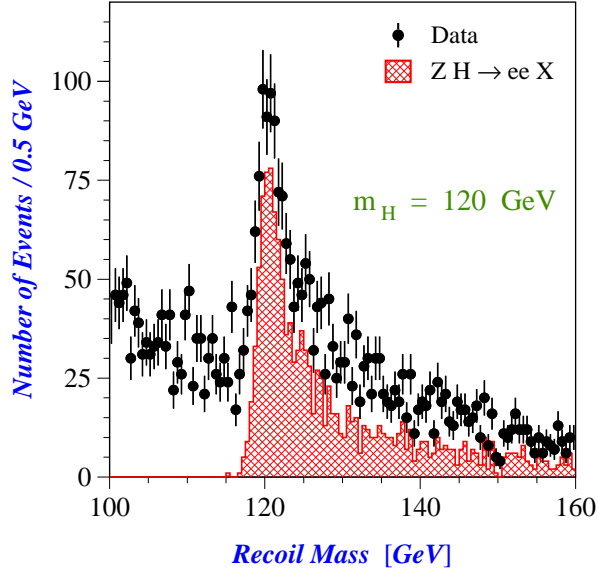}}
\caption{\label{fig2}\it Production of Higgs particles in the Higgs--strahlung channel
$e^+e^- \ra ZH \ra e^+e^- X$; Ref.\cite{34a}.}
\end{figure}

The recoiling $Z$ boson in the two--body reaction $\epem \ra ZH$ is
mono--energetic
and the mass can be derived from the energy of the $Z$ boson,
$M_H^2 =s -2\sqrt{s} E_Z +M_Z^2$.
Initial state bremsstrahlung and beamstrahlung
smear out the peak slightly,
as shown in Fig.~\ref{fig2}.
A similarly clear peak can be observed in the fusion process
\ee $\ra \overline{\nu}_e \nu_e H$
by collecting the decay products of the
Higgs boson. 

\paragraph{\underline{Mass and Width:}}
The mass of the Higgs boson can be measured very accurately by analyzing
the recoil $Z$ spectrum in Higgs--strahlung events.
Experimental simulations \cite{40} have demonstrated that the error in the mass measurement
can be reduced to
\begin{eqnarray*}
\delta M_H \lesssim 50\,\makebox{MeV}
\end{eqnarray*}
in high--luminosity runs.

The width of the \SM Higgs boson can be determined in an 
almost completely model--independent way in the
difficult intermediate mass range in which the Breit--Wigner
form cannot be reconstructed at an \ee collider.
Measuring the branching ratio $BR_i$ in the decay and the
partial width $\Gamma_i$ in the production process, the total width
$\Gamma_H$ can be derived from
\begin{eqnarray*}
\Gamma_H = \Gamma_i / BR_i 
\end{eqnarray*}
The two channels $i=WW$ \cite{40a1,40a2} 
and $i=\gamma\gamma$ \cite{40b} 
are useful for this analysis \cite{40c}.
The partial width $\Gamma_{WW}$ can be extracted from the size of the $WW$ 
fusion cross section \cite{40d}
while $\Gamma_{\gamma\gamma}$ can be measured in
the Compton collider mode \cite{40e}. 
The accuracies of a few percent match the expected accuracy in scanning the
Breit--Wigner excitation at a muon--collider.

\paragraph{\underline{Spin and Parity:}}
The angular distribution of the $Z/H$ bosons in the Higgs--strahlung
process is sensitive to the external quantum numbers
of the Higgs particle \cite{33}.
Since the amplitude is given by
${\cal A} (0^+) \sim \varepsilon_{Z^{\ast}} \cdot \varepsilon^{\ast}_Z$,
the $Z$ boson is produced in a state of longitudinal polarization
at high energies.
As a result, the angular distribution
${\rm d}\sigma/{\rm d}\cos\theta  \sim \lambda
\sin^2 \theta + 8M_Z^2/s$
approaches the spin--zero law
$ \sin^2\theta$ asymptotically.
This may be contrasted with the distribution
$\sim 1 + \cos^2 \theta$
for negative parity states, which follows from the transverse
polarization amplitude ${\cal A} (0^-) \sim \varepsilon_{Z^{\ast}}
\times \varepsilon^{\ast}_Z \cdot k_Z$.
It is also characteristically different from the background process
\ee $\ra ZZ$ which is strongly peaked in the forward/backward
direction.

\begin{figure}[t]
\centerline{\includegraphics[width=12cm]{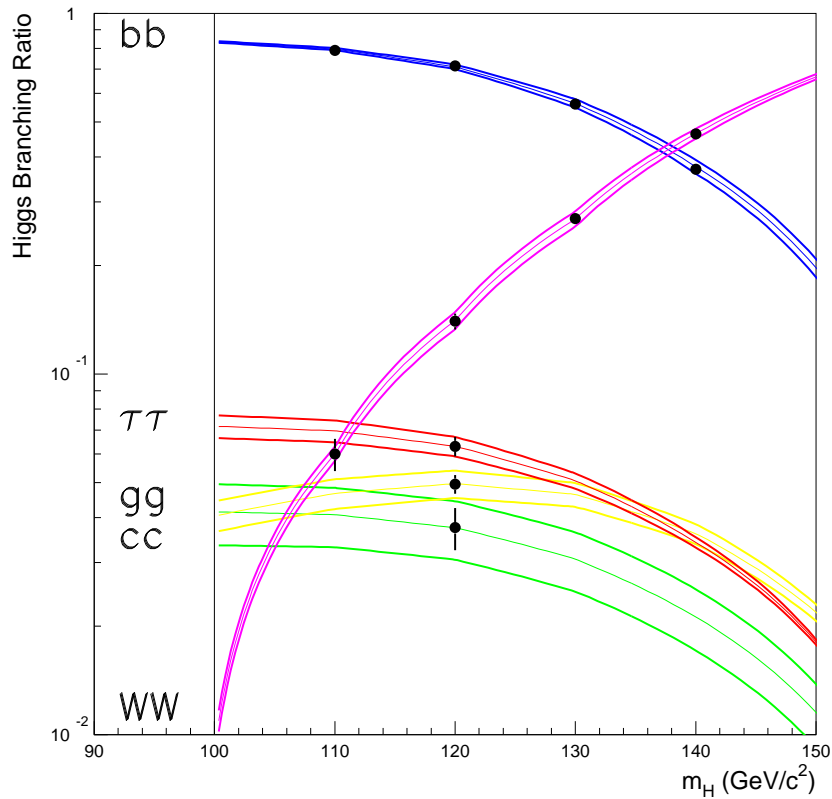}}
\caption{\label{fig3}\it Branching ratios of \SM Higgs decays into fermion and $WW^*$ pairs; 
Ref.\cite{40a2}.}
\end{figure}

\paragraph{\underline{Higgs Couplings:}}
Since the fundamental \SM particles acquire masses by
means of  the interaction
with the Higgs field, the scale of the
Higgs couplings to fermions and gauge bosons is
set by the masses of the particles:
\begin{eqnarray*}
g_{HVV} = 2 [ \sqrt{2} G_F ]^{1/2} M_V^2 
\qquad \mbox{and} \qquad
g_{Hff} = [ \sqrt{2} G_F ]^{1/2} M_f
\end{eqnarray*}
It will be a very
important task to measure the Higgs couplings to the
fundamental particles \cite{40a1,40a2}
since they are uniquely predicted by the very nature
of the Higgs mechanism.
The Higgs couplings to massive gauge bosons can be
determined from the measurement of the production cross sections
with an accuracy of $\pm 1\%$, the $HZZ$ coupling in the
Higgs--strahlung and the $HWW$ coupling in the fusion process. For
Higgs couplings to fermions, either loop effects in
$H \rightleftharpoons gg, \gamma \gamma$
[mediated by top quarks] can be exploited, or the direct measurement of
branching ratios
$H \ra b \overline{b}, c \overline{c}, \tau^+ \tau^-, gg$
in the lower part of the intermediate mass
range. 
This is exemplified in Fig.~\ref{fig3}.
For $M_H = 120$\,GeV the following accuracy $\delta BR / BR$
can be achieved \cite{40a2}
in the determination of the Higgs decay branching ratios:
\begin{eqnarray*}
bb: 2.4\% & \qquad &           WW^*\!: 5.4\% \\
cc: 8.3\% & \qquad & \quad\,\, gg    : 5.5\% \\
\tau\tau: 6.0\% & 
\end{eqnarray*}
By measuring the ratio of the $\tau\tau$ to the $bb$ branching ratios
\begin{eqnarray*}
\frac{BR(H\ra \tau\tau)}{BR(H\ra bb)} = \frac{m_\tau^2}{3m_b^2(M_H)}
\end{eqnarray*}
the linear dependence of the Yukawa couplings on the fermion masses
can be tested very nicely.
A direct way to determine the Yukawa coupling of the intermediate
mass Higgs boson to the top quark in the range $m_H \le 120$\,GeV is
provided by the bremsstrahlung process $e^+e^- \ra t\overline{t} H$
in high energy $e^+e^-$ colliders \cite{40f,40ff}.
The absolute values of the Yukawa couplings can be reconstructed by
combining decay branching ratios with the production cross sections.

\paragraph{\underline{Higgs Self--couplings:}}
To generate a non--zero value of the Higgs field in the vacuum,
the minimum of the Higgs potential must be shifted away from the origin.
Rewriting the potential
\begin{eqnarray*}
V & = & \lambda \, [ \mid\!\varphi\!\mid^2 - 
{\textstyle \frac{1}{2}}\,v^2 ] ^2 \\[0.1cm]
  & = & \frac{M_H^2}{2}\, H^2 + \frac{M_H^2}{2 v}\, H^3 + 
\frac{M_H^2}{8 v^2}\, H^4 
\end{eqnarray*}
in terms of the physical Higgs field $H$, the potential can be
reconstructed by measuring the trilinear and quartic couplings.
At a high--luminosity \ee collider, the trilinear coupling can be
tested in the double Higgs--strahlung process:
\begin{eqnarray*}
e^+ e^- \ra Z + HH
\end{eqnarray*}
The splitting of a virtual Higgs boson into two real Higgs bosons
is determined by the trilinear Higgs coupling:
$e^+e^- \ra Z + H^* \, [\ra HH]$.
Even though the cross section is less than 1\,fb \cite{40g}, the coupling
can nevertheless 
 be measured with an accuracy better than 20\%, cf.~Ref.~\cite{40h}.
Thus an essential element of the mechanism responsible
for the spontaneous symmetry breaking in the scalar sector can be
established experimentally at the high--luminosity collider.

\section[]{Supersymmetry}

Even though there is no direct experimental evidence so far for the
realization of supersymmetry in Nature, this concept has so many
attractive features that it can be considered as a  prime
target of present and future
experimental particle research \cite{40l}.
Arguments in favor of supersymmetry are deeply
rooted in particle physics. 
Supersymmetry may play an important
r$\hat{\rm o}$le in a quantum theory of gravity. In relating particles
of different spins to each other, i.e. fermions and bosons, low--energy
supersymmetry  stabilizes the masses
of fundamental Higgs scalars in the background of very high energy
scales associated with grand unification. Besides solving this
hierarchy problem, supersymmetry may even be closely related to the
physical origin of the Higgs phenomenon itself: In a
supergravity inspired GUT
realization with universal scalar masses at the GUT scale, the
evolution of one of the scalar masses squared 
down to the electroweak scale can become negative and
can thus give rise to spontaneous
symmetry breaking if the top mass has a value between about 100 to 200
GeV while all other squared masses  of squarks and sleptons remain
positive so that ${\rm U(1)_{EM}}$ and ${\rm SU(3)_C}$ remain unbroken.\\

The minimal supersymmetric extension of the Standard Model (MSSM)
is based on
the \SM\ group ${\rm SU(3)\times SU(2) \times U(1)}$. 
The 
\MSSM incorporates a  spectrum of five Higgs
particles (representative for a wide class of models).
At tree level,
the mass of the lightest Higgs boson $h^0$ is smaller than the $Z$ mass;
 the bound is  shifted to $\stackrel{<}{\sim} 140$
 GeV by radiative corrections.
The masses of the heavy neutral
and charged Higgs bosons can be expected in the
range of the electroweak symmetry breaking scale.
The gauginos are the
supersymmetric spin--$\frac{1}{2}$ partners of the gauge bosons. The
quark and lepton matter particles are associated with scalar
supersymmetric particles, squarks and sleptons. To preserve
supersymmetry, two Higgs doublets are needed to endow down as well
 as up--type fermions with masses; the supersymmetric
partners of the Higgs bosons are spin--$\frac{1}{2}$ higgsinos.
[Charged/neutral higgsinos mix in general with the non--colored
gauginos, forming charginos and neutralinos.]     Supersymmetric
partners carry a multiplicative quantum number $R=-1$ ($R=+1$ for
ordinary particles) which is conserved in this model. Supersymmetric
particles are therefore generated in pairs and the lightest
supersymmetric particle  $LSP$  is stable.

Strong support for supersymmetry and the \MSSM particle
spectrum in the mass range of several hundred GeV
follows from the high--precision
measurement of the electroweak
mixing angle $\sin^2 \theta_W$  \cite{41}. The value
predicted by the \MSSM and the value determined by the LEP
and other experiments,
\begin{eqnarray*}
MSSM:                \makebox[5mm]{} \sin^2 \theta_W & = 0.2336 \pm 0.0017 \\
EXP \makebox[4mm]{}: \makebox[5mm]{} \sin^2 \theta_W & = 0.2316 \pm 0.0002
\end{eqnarray*}
match surprisingly well, the difference being less than about 2\,per--mille.\\

A central problem of supersymmetric theories is the breaking mechanism.
Several scenarios have been proposed and experimental consequences
have been elaborated in a few cases: 
gravity mediated supersymmetry breaking mSUGRA \cite{40m};
gauge mediated supersymmetry breaking GMSB \cite{40n};
anomaly mediated supersymetry breaking AMSB \cite{40p};
Scherk--Schwarz supersymmetry breaking SSSB \cite{40q}.
Mass spectra of the supersymmetric particles are quite different in these scenarios
so that high--precision measurements of the particle properties will shed light
experimentally on this theoretical problem.
Moreover, extrapolations can be performed in a stable way
which will allow to reconstruct the basic supersymmetric theory
eventually at a scale close to the Planck scale.
Precision experiments at a high--luminosity \ee linear collider
are therefore
 expected to advance the understanding of supersymmetry in an essential way.

\subsection[]{\SUSY Higgs Particles}

The Higgs spectrum in the minimal supersymmetric extension of the
Standard Model \cite{40i} consists of five states: $h^0, H^0,
A^0$ and $H^\pm$. Besides the masses, two mixing angles
define the properties of the scalar particles
and their interactions with gauge bosons, fermions
 and the self--interactions: the ratio of
the two vacuum expectation values $\tb = v_2/v_1$ and a mixing
angle $\alpha$ in the neutral ${\cal {CP}}$--even sector. 
Supersymmetry leads to several relations among 
the masses and mixing angles
and, in fact, only
two parameters are independent.

\paragraph{\underline{Neutral Higgs Bosons:}}
The lightest neutral Higgs boson
will decay mainly into fermion
pairs since its mass is smaller than $\sim$ 140 GeV.
This is also the dominant
decay mode of the pseudoscalar boson $A^0$.
For values of $\tb$ larger than unity and for masses less than
$\sim$ 140 GeV, the     main decay modes of the neutral Higgs bosons are
decays into $b \bar{b}$ and $\tau^+ \tau^-$     pairs;  the
branching ratios
are of order    $ \sim 90\%$ and $8\%$, respectively.
The decays
into $c \bar{c}$ pairs and gluons [proceeding through $t$ and $b$ quark
loops] are  suppressed,  for large $\tb$ strongly. For
large masses, the top decay channels $H^0, A^0 \rightarrow t\bar{t}$
open up; yet
this mode remains suppressed for large $\tb$. For large $\tb$,
the neutral
Higgs bosons decay almost universally into $\bbb$ and $\tau^+\tau^-$
pairs.
If the mass is high enough, the heavy ${\cal {CP}}$--even Higgs boson
$H^0$  can in
principle decay into weak gauge bosons, $H^0 \rightarrow WW, ZZ$.
Since the partial widths are proportional to $\cos^2 (\beta-\alpha)$,
they are
strongly suppressed and the gold--plated $ZZ$ signal of the heavy
\SM Higgs
boson is lost in the supersymmetric extension.
The heavy neutral Higgs boson $H^0$ can also decay into two lighter
Higgs bosons. These modes, however, are restricted to small
domains in the parameter space.
Other possible channels are decays into supersymmetric particles.
While sfermions are likely too heavy to affect Higgs decays in the
mass range
considered here, Higgs boson decays into charginos and neutralinos could
eventually be important since the masses of
some of these particles are expected to be of
order $M_Z$. 
[These  channels are   summarized in Ref.\cite{40k}.]
The charged Higgs particles decay into fermions but also,
if    allowed   kinematically, into the lightest neutral Higgs
boson and a
$W$ boson. Below the $tb$       and $Wh$ thresholds, the charged Higgs
particles will decay mostly into $\tau \overline{\nu}_\tau$ and $c\bar{s}$
pairs, the former being dominant for $\tb >1$.  For
large $M_{H^\pm}$ values, the top--bottom decay mode
$H^+ \rightarrow      t\bar{b}$ becomes dominant.

Adding up the various decay modes, the width of all five Higgs bosons
remains very narrow, being of order 10 GeV even for large masses.\\

The search for the neutral \SUSY Higgs bosons at
$\epem$ colliders will be a straightforward extension
 of the search  performed
at LEP2.  This collider is expected to cover the mass range
up to $\sim  110$\,GeV 
for neutral Higgs bosons, depending on $\tb$.
Higher energies, $\sqrt{s} \sim 250$ GeV,
are required to sweep the entire parameter space of the \MSSM.
The main production mechanisms of
 neutral Higgs bosons at $\epem$
colliders \cite{40k,40i} are the Higgs--strahlung process
and associated pair production,
\begin{eqnarray*}
(a) \quad {\rm Higgs-strahlung}: \quad    \epem & \ra & Z+h/H \\
 (b) \quad \, {\rm Pair \ production}  \makebox[3.75mm]{}: \makebox[4mm]{}
  \epem & \ra & A+h/H
\end{eqnarray*}
as well as the related fusion processes.
The ${\cal {CP}}$--odd Higgs boson $A^0$ cannot be produced in
fusion processes because it does not couple to
gauge bosons in  leading order.

The cross sections for the four Higgs--strahlung and pair production
processes can be expressed as
\begin{eqnarray*}
\sigma(\epem \ra Zh) & =& \sin^2(\beta-\alpha) \ \sigma_{SM}                    \\
\sigma(\epem \ra ZH) & =& \cos^2(\beta-\alpha) \ \sigma_{SM}
\end{eqnarray*}
and
\begin{eqnarray*}
\sigma(\epem \ra Ah) & =& \cos^2(\beta-\alpha) \ \bar{\lambda} \ \sigma_{SM}    \\
\sigma(\epem \ra AH) & =& \sin^2(\beta-\alpha) \ \bar{\lambda} \ \sigma_{SM}
\end{eqnarray*}
where $\sigma_{SM}$ is the \SM cross section for
Higgs--strahlung and
the factor $\bar{\lambda}$ accounts for the suppression of the
$P$--wave cross sections near the threshold. The cross sections for the
Higgs--strahlung and for the pair production as well as the cross
sections for the production of the light and the heavy neutral
Higgs bosons $h^0$ and $H^0$ are
mutually complementary to each other, coming either with a coefficient
$\sin^2(\beta-\alpha)$ or $\cos^2(\beta-\alpha)$. As a result,
since $\sigma_{SM}$ is large,
at least the lightest ${\cal {CP}}$--even Higgs boson must be detected.

The cross section for $hZ$ in the Higgs--strahlung process
is large for values of $M_h$ near the upper bound. 
The heavy ${\cal {CP}}$--even and ${\cal {CP}}$--odd Higgs bosons $H$ and $A$,
on the other hand, are produced pairwise in this limit:
$e^+ e^- \ra A H$.
The decoupling limit becomes effective for 
 heavy Higgs masses above 250 to 300\,GeV.
The discovery limit is therefore set by the beam energy independently
of the mixing parameter $\tb$,
in contrast to LHC where the heavy Higgs bosons cannot be detected
individually in parts of the parameter space.

\paragraph{\underline{Charged Higgs Bosons:}}
The charged Higgs bosons,
if lighter than the top quark, can be
produced in top decays, $t \ra b + H^+$, with a branching ratio
varying between $2\%$ and $20\%$ in the kinematically allowed region.
                                 Since for $\tb$ larger than
unity, the charged Higgs bosons will decay mainly into
$\tau \nu_\tau$,
this results in a surplus of $\tau$ final states over $e, \mu$ final
states in $t$ decays, an apparent breaking of lepton
universality. For large Higgs masses the dominant decay
mode is the top decay $H^+ \ra t \overline{b}$.
In this case the charged Higgs particles must be pair produced
in \ee colliders:
\[
              \epem \ra H^+H^-
\]
\noindent
The cross section depends only on the charged Higgs mass. For
small Higgs masses the cross
section is of order 100 fb at $\sqrt{s} = 500$ GeV,
but it drops very quickly due to the
$P$--wave suppression $\sim \beta^3$ near the threshold.\\

\noindent
{\bf SUMMARY:}
The preceding discussion of the \MSSM Higgs sector at $\epem$
linear colliders can be summarized in the following two points:

\noindent
{$(i)$} The lightest ${\cal {CP}}$--even Higgs particle $h^0$ can be
detected in the entire range of the \MSSM parameter space, either
in      the Higgs--strahlung process $\epem \rightarrow hZ$
or in      pair production
$\epem \rightarrow hA$ \cite{40}.
In fact, this conclusion holds true
even at a c.m. energy of 250~GeV, even 
if invisible neutralino decays are allowed for.

\noindent
{$(ii)$}  The area in the parameter
space where {\it all} \SUSY Higgs bosons can be discovered individually at
\ee colliders, is characterized by $M_A \stackrel{<}{\sim}
\frac{1}{2} \sqrt{s}$, independently of $\tb$.
The $H^0, A^0$ Higgs bosons can be produced either
in Higgs--strahlung or in $Ah, AH$ associated production;
charged Higgs bosons will be produced in $H^+H^-$ pairs
up to the kinematical limit.

\subsection[]{Supersymmetric Particles}

\paragraph{\underline{Charginos and Neutralinos}}
The two charginos $\cc_i$ and the four neutralinos $\cn_i$, mixtures of
the [non--colored] gauginos and higgsinos, are expected to be the
lightest
supersymmetric particles.
In the \MSSM with conserved $R$--parity,
the neutralino $\cn_1$ with the smallest mass, assumed to be the
lightest   supersymmetric particle, is stable. The heavier
neutralinos and the
charginos decay into (possibly virtual) gauge and Higgs bosons plus
the $LSP$,
$\cn_i \lra  \cn_1 + Z$ and $\cn_1 + W$,
or if they are heavy enough, into neutralino/chargino cascades,
and leptons plus sleptons \cite{42}.

Neutralinos and charginos are  easy to detect at $\epem$
colliders.
They are produced in pairs
\beq
\epem & \ra & \cc_i  \wt{\chi}^-_j \hspace*{1cm} [i,j=1,2] \non \\
\epem & \ra & \cn_i  \cn_j \hspace*{1.2cm} [i,j=1,..,4]
\eeq
through $s$--channel $\gamma, Z$ exchange and $t$--channel selectron or
sneutrino exchange.
Since the cross sections are as large
as ${\cal O}$(100~fb), enough events will be produced to discover these
particles for masses nearly up to the kinematical limit.

\begin{center}
\begin{figure}[t]
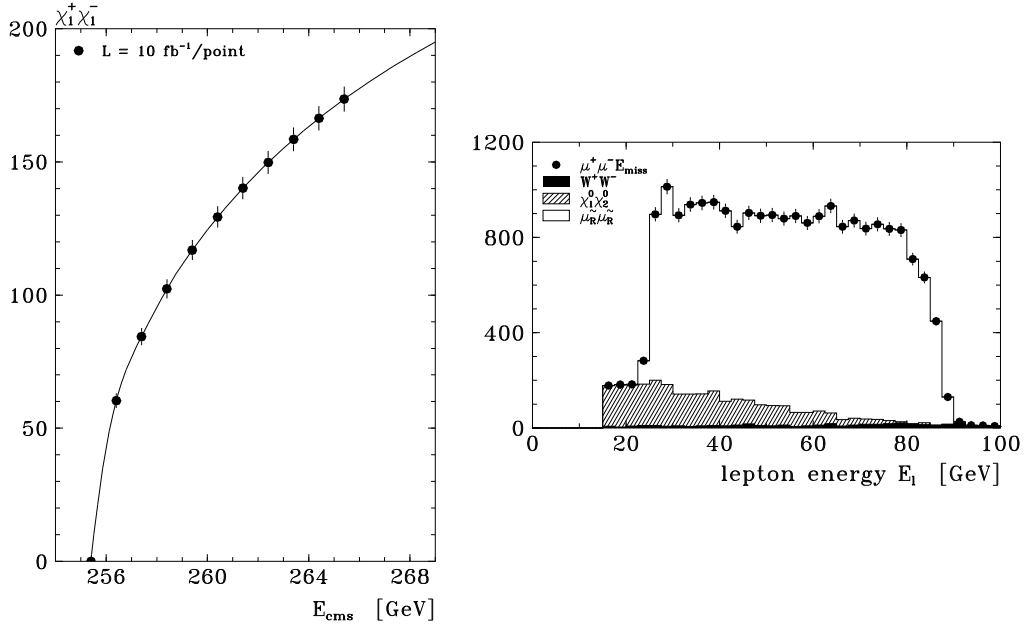

\unitlength1cm
\begin{picture}(15,10)
\put(.7,0){\includegraphics[width=10cm,angle=90]{c11scan.eps}}
\put(7.1,2){\includegraphics[width=6cm,angle=90]{mur132.emu.w320.eps}}
\end{picture}
\caption{\label{fig4}\it The excitation curve for chargino production 
$e^+e^- \ra \tilde{\chi}_1^+ \tilde{\chi}_1^-$ near the threshold and
energy distribution of the final state $\mu$ in the decay 
$\tilde{\mu}_R \ra \mu + \tilde{\chi}_1^0$ in flight; Ref.\cite{42a}.}
\end{figure}
\end{center}

The properties of the charginos and neutralinos can be studied in great
detail at \ee colliders. From the fast onset
$\sim \beta$ of the spin $-\frac{1}{2}$
excitation curve near the
threshold, the masses can be measured very accurately 
within less than 100 MeV,  Fig.~\ref{fig4}.
Using polarized $e^\pm$ beams, the
decomposition of the states,  $\cc_i = \alpha \tilde{W}^+
+ \beta \tilde{H}^+$
into wino and higgsino components can be determined \cite{43a} since left--handed
electrons couple to sneutrinos in the $t$--channel but right--handed
electrons do not, so that the
energy and angular dependence of the cross
sections is different for the two polarization states \cite{44}.
In a similar way the properties of neutralinos can be explored \cite{44a}.

\paragraph{\underline{Sleptons and Squarks:}}
The superpartners of the right--handed leptons decay into the
associated \SM partners and neutralinos/charginos. In major parts of
the \SUSY parameter space  the dominant decay mode is
$ \tilde{\mu}_R \ra \mu + \cn_1$ \cite{42}.
For the superpartners of the left--handed sleptons, the decay pattern
is slightly more complicated since, besides the $\tilde{\chi}_1^0$
channels,
decays into leptons and charginos are also possible.
In $\epem$ and $e^- e^-$ collisions, sleptons are produced in pairs:
\beq
\epem &\lra & \tilde{\mu}_L^+\tilde{\mu}_L^- \ , \
 \tilde{\mu}_R^+\tilde{\mu}_R^-
   \ , \
\tilde{\tau}_L^+\tilde{\tau}_L^- \ , \ \tilde{\tau}_R^+\tilde{\tau}_R^-
\non \\
\epem &\lra & \tilde{\nu}_L \overline{\tilde{\nu}}_L \non \\
\epem &\lra & \tilde{e}_L^+\tilde{e}_L^- \ , \
\tilde{e}_R^+\tilde{e}_R^- \ , \
\tilde{e}_L^+\tilde{e}_R^- \ , \ \tilde{e}_R^+\tilde{e}_L^- \non \\
e^-e^- &\lra & \tilde{e}_L^-\tilde{e}_L^- \ , \
\tilde{e}_R^-\tilde{e}_R^- \ , \
\tilde{e}_L^-\tilde{e}_R^- 
\eeq
For charged sleptons, the production proceeds via $\gamma,Z$ exchange
in the $s$--channel, in the case of selectrons, also by additional
$t$--channel   neutralino exchange.
For sneutrinos, the
process is mediated by $s$--channel $Z$--exchange and, in the case of
electron--sneutrinos, also by the $t$--channel exchange of charginos.

The cross sections for the pair production of sleptons are of the order
of $10^{-1}$ to $10^{-2}$\,pb so that
their discovery is very easy up to the kinematical limit.
From the $P$--wave onset $\sim \beta^3$ of the annihilation cross section
the masses can in general be determined \cite{42a}
at a level of 200 to 300\,MeV; 
the sharper onset of selection production in $e^-e^-$ scattering \cite{44b}
will reduce this number further.
Enough events will be produced to study
their detailed properties.
The polarization of the $e^\pm$ beams will
help  to identify the couplings of these particles. 

The endpoints in the decay spectra of $\tilde{\mu}_R \rightarrow
\mu + \tilde{\chi}^0_1$ can be exploited to determine the LSP
mass with an accuracy of 100 MeV, cf. Fig.~\ref{fig4}.

If one of the stop states is light enough due to the strong $LR$ Yukawa
mixing, these particles may be pair produced even
at a 500\,GeV collider:
\begin{eqnarray*}
e^+e^- \ra \tilde{t_i}  \overline{\tilde{t_j}} \qquad i,j = 1,2
\end{eqnarray*}
By measuring the $LR$ asymmetry of the production cross sections, the
$\tilde{t}_L/\tilde{t}_R$ mixing angle can be determined
to high accuracy $\delta\cos\theta_{\tilde{t}} \approx 0.01$ \cite{44c}.

\subsection[]{Supersymmetry Breaking}

\renewcommand{\thefootnote}{\arabic{footnote}}\addtocounter{footnote}{-1}

The high precision with which masses, couplings and mixing
parameters will be determined at \ee colliders, can be exploited
to test the 
mechanism for supersymmetry breaking and the
structure of the underlying theory.
In minimal supergravity mSUGRA the breaking of supersymmetry
is mediated by gravity from a hidden sector to the eigenworld,
generating soft \SUSY breaking parameters at the grand unification 
scale\footnote{
If gravitational interactions would become strong not at a
very high scale but near the electroweak scale \cite{44e1}, collider
experiments could probe the additional spatial dimensions through
which gravitational fields would propagate \cite{44e2}.
Contact interactions and missing energy events could signal Planck scales in
4\,+\,n dimensions up to about 10\,TeV.
Thus, the basic space--time structure can be explored in these experiments.
}.
The parameters are generally assumed to be universal at that scale
in the gaugino and the scalar sectors.
In gauge mediated supersymmetry breaking, gauge interactions connect the
mechanism to the eigenworld at a scale possibly between 10 and $10^3$\,TeV.
Mass spectra in mSUGRA and GMSB are characteristically different, 
the splitting between sleptons and squarks being larger in GMSB.
Moreover, since the gravitino in GMSB is very light, $\tilde{\chi}_1^0$ or $\tilde{\tau_1}$
may be long lived giving rise to displaced photons or stable heavy tracks in the decays 
$\tilde{\chi}_1^0 \ra \gamma \tilde{G}$ or $\tilde{\tau_1} \ra \tau \tilde{G}$ 
with the gravitino $\tilde{G}$ escaping undetected \cite{44d}.\\

If minimal supergravity is the underlying theory, the observable
properties of the superparticles can be expressed by a small
set of parameters defined at the GUT scale. 
Basically five
parameters specify the properties of the particles in
 the supersymmetric sector. The scalar mass
parameter $m_0$, the ${\rm SU(2)}$ gaugino mass $M_{1/2}$,
the trilinear coupling $A_0$ and the sign of
the coupling $\mu$ of
the Higgs doublets in the superpotential, and $\tan \beta$, the ratio
of the vacuum expectation values $v_2/v_1$. 
 Evolving the scalar masses from the
GUT scale down to low energies, it turns out that non--colored
particles are in general
significantly lighter than colored particles.
The lightest of the non--colored gauginos/higgsinos
and sleptons could have masses in the range of 100 to 200\,GeV.
Since only a few parameters determine the low energy theory of
the evolution from the GUT scale down to the electroweak scale,
many relations can be found among the masses of the superparticles
which can stringently be tested at \ee colliders \cite{44}. 
Two examples should illustrate the 
potential of \ee facilities in this context.\\[3mm]
\noindent
{$(i)$} The \underline{\it gaugino masses} at the
scale of  ${\rm SU_2 \times U_1}$ symmetry breaking are
related to the common ${\rm SU_2}$ gaugino mass $M_{1/2}$ at the
GUT scale by the running gauge couplings:
\beq
M_i = \frac{\alpha_i}{\alpha_{GUT}} M_{1/2}  \makebox[5mm]{}
i = SU_3, SU_2, U_1 \non
\eeq
with $\alpha_{GUT}$ being the gauge coupling at the unification
scale. The mass relation in the non--color sector
\beq
\frac{M_1}{M_2} = \frac{5}{3} \tan^2 \theta_W \approx \frac{1}{2}  \non
\eeq
can be tested very well by measuring the masses and production
cross sections of charginos/neutralinos and sleptons.\\[3mm]
\noindent
{$(ii)$} In a similar way the \underline{\it slepton masses} can be
expressed in terms of a common scalar mass parameter $m_0$ at
the GUT scale, contributions $\sim M_{1/2}$ due to the
evolution from the GUT scale down to low energies, and the
D terms related to the electroweak symmetry breaking.
These expressions give rise to relations among the slepton masses:
\beq 
m^2  (\tilde{l}_L) - m^2 (\tilde{\nu}) & = & - (1 -2 \sin^2 \theta_W)
\cos2 \beta m^2_Z \\  \non
m^2  (\tilde{l}_L) - m^2 (\tilde{l}_R) & = & \kappa\, M^2_{1/2} -
{\scriptstyle \frac{1}{2}}
(1 - 4 \sin^2 \theta_W) \cos 2 \beta m^2_Z   \non
\eeq
with $\kappa=0.37$.
The second relation follows from the hypothesis that
the scalar masses are universal at the GUT scale,
 in particular  $m^2 (5^{\ast}) = m^2 (10)$
within ${\rm SU_5}$. This assumption can be tested by relating
the mass difference between $\tilde{e}_L$ and $\tilde{e}_R$ to the
${\rm SU_2}$ gaugino mass.\\[2mm]
The typical result of an overall fit to the fundamental
mSUGRA parameters at the GUT scale and tg$\beta$ is illustrated in 
Table\,\ref{tab2}.\\[2mm]

\begin{center}
\begin{figure}[t]
\epsfig{bbllx=20,bblly=410,bburx=535,bbury=655,file=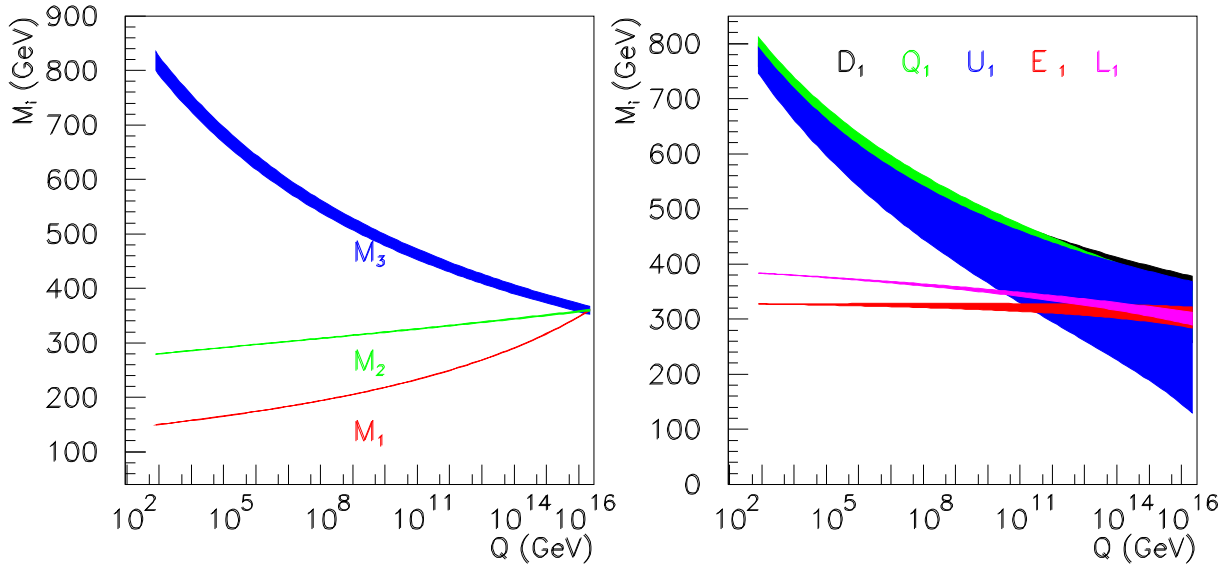,width=16.5cm}
\caption{\label{fig5}\it Example for the reconstruction of the fundamental
supersymmetric mass parameters at the unification scale by 
evolution from the electroweak scale in the gaugino and
scalar sectors; for details see Ref.\cite{44f}.}
\end{figure}
\end{center}

In a procedure which reveals more clearly the structure of the underlying 
theory\footnote{This procedure is particularly important
for non-universal theories.},  the parameters may not only be fitted
by assuming a universal set at the GUT scale from the start, but
the set itself may be reconstructed by evolving the mass parameters 
from the electroweak scale to the unification scale \cite{44f}, 
cf. Fig.~\ref{fig5}. For gauginos and sleptons the
reconstruction of the universal mass parameters is excellent.
Due to mutual cancelations it is much more
difficult for the squark and Higgs sectors. Nevertheless,
this is the only way to reconstruct operationally the
fundamental supersymmetric theory near the Planck scale
from experimental observations at the electroweak scale.\\

\noindent
Precision tests of supersymmetric particles in \ee collider experiments
can thus open a window to energy scales close to the Planck
scale where gravity, the fourth of the fundamental forces, becomes
an integral part of physics.

\begin{table}
\begin{center}
\begin{tabular}{|c||c|c|}
\hline
\rule[-3mm]{0mm}{9mm} Parameter      &  Theor. Value   &  Meas. Error  \\
\hline
\rule[-0mm]{0mm}{6mm} $m_0$          &  100\,GeV       &  0.09\,GeV    \\
                      $M_{1/2}$      &  200\,GeV       &  0.10\,GeV    \\
                      $A_0$          &  0\,GeV         &  6.30\,GeV    \\
\rule[-4mm]{0mm}{0mm} $\tan\beta$    &  3              &  0.02         \\
\hline
\end{tabular}
\caption{\label{tab2}{\it Measurement of universal mSUGRA parameters;
Ref.~\cite{42a}.}}
\end{center}
\end{table}


\end{document}